\documentstyle[twoside,fleqn,espcrc2,epsfig]{article}

\newcommand{\be}{\begin{equation}}
\newcommand{\ee}{\end{equation}}
\newcommand{\bea}{\begin{eqnarray}}
\newcommand{\eea}{\end{eqnarray}}

\newcommand{\AmS}{{\protect\the\textfont2
  A\kern-.1667em\lower.5ex\hbox{M}\kern-.125emS}}
\def\lapproxeq{\lower .7ex\hbox{$\;\stackrel{\textstyle                                                   
<}{\sim}\;$}}                                                   
\def\gapproxeq{\lower .7ex\hbox{$\;\stackrel{\textstyle                                                   
>}{\sim}\;$}}                                                  
\title{$F_2$ at low $Q^2$}

\author{A.D. Martin \address{Department of Physics,
 University of Durham,Durham, DH1 3LE, UK.}%
        ,M.G. Ryskin \address{Petersburg Nuclear Physics Institute, 188350,                                                   
Gatchina, St. Petersburg, Russia. }
	and
	A.M. Stasto \address{H.\ Niewodniczanski Institute of Nuclear Physics,                                                   
31-342 Krakow, Poland. }}
\begin{document}

\begin{abstract}
We analyse the data for the proton structure function $F_2$ over the entire $Q^2$                                        
domain, including especially low $Q^2$, in terms of perturbative and                                                   
non-perturbative QCD contributions.  The small distance                                                   
configurations are given by perturbative QCD, while the large                                                   
distance contributions are given by the vector dominance model                                                   
and, for the higher mass $q \overline{q}$ states, by the additive                                                   
quark approach.
\end{abstract}

\maketitle

\section{INTRODUCTION}
                                                   
There now exist high precision deep inelastic $ep$ scattering                                                   
data \cite{H1,ZEUS} covering both the low $Q^2$ and high $Q^2$                                                   
domains, as well as measurements of the photoproduction cross                                                   
section.  The interesting structure of these measurements, in                                                   
particular the change in the behaviour of the cross section with $Q^2$ at $Q^2 \sim                                        
0.2 {\rm GeV}^2$, highlight the importance of obtaining a theoretical QCD                                                   
description which smoothly links the non-perturbative and perturbative domains.                                                   
                                                   
In any QCD description of a $\gamma^* p$ collision, the first                                                   
step is the conversion of the initial photon into a $q                                                   
\overline{q}$ pair, which is then followed by the interaction of                                                   
the pair with the target proton, see Fig.1.  Let $\sigma (s, Q^2)$ be the                                                   
total cross section for the process $\gamma^* p \rightarrow X$                                                   
where $Q^2$ is the virtuality of the photon and $\sqrt{s}$ is the      
$\gamma^* p$ centre-of-mass energy.
We can write the dispersion relation in the following way:
\be                                                   
\label{eq:a1}                                                   
{\scriptstyle \sigma (s, Q^2) \; = \; \sum_q \: \int_0^\infty \:
 \frac{d M^2}{(M^2 +                                                   
Q^2)^2} \; \rho (s, M^2) \: \sigma_{q \overline{q} + p} (s, M^2)}
\ee                                                   
where the spectral function $\rho (s, M^2)$ is the density of $q                                                   
\overline{q}$ states.                                                   
Following \cite{BK} we may divide the                                          integral into                                                   
two parts: the region $M^2 < Q_0^2$                                                    
described by the vector meson                                                   
dominance model (VDM) and the region $M^2 > Q_0^2$ described by                                                   
perturbative QCD.
 To exploit further this idea we must achieve a better separation between                                                   
the short and long distance contributions.  To do this we take a                                                   
two-dimensional integral over the longitudinal ($z$), and transverse                                                   
momentum ($k_T$) components of the quark, see Fig1.
                                                   
The contribution coming from the small mass region is pure VDM.  The                                                   
part which comes from large $k_T$ of the quark can be calculated                                                   
by perturbative QCD in terms of the known parton distributions,                                                   
whereas for small $k_T$ we will use the additive quark model and                                                   
the impulse approximation.  That is only one quark interacts with                                                   
the target and the quark-proton cross section is well                                                   
approximated by one third of the proton-proton cross section.                                            

\section{The $\gamma^* p$ cross section}
                   
The spectral function $\rho$ occurring in (\ref{eq:a1}) may be                                                   
expressed in terms of  the $\gamma^* \rightarrow q \overline{q}$                                                   
matrix element $\cal{M}$.  We have $\rho \: \propto | {\cal M} |^2$
and in terms of the quark momentum variables $z$ and  $k_T^2$ 
the cross section
in equation (\ref{eq:a1}) becomes
\hspace{-0.5cm}
\bea
{\scriptstyle
\sigma_T = \sum_q \alpha \frac{e_q^2}{4 \pi^2} \sum_{\lambda \: =
 \: \pm \: 1} \int dz  d^2 k_T  ({\cal M}_T {\cal M}_T^*)
N_c   \frac{1}{s}  {\rm Im}  A_{q \overline{q} + p}} 
\nonumber
\eea                          
\be
\label{eq:a2}
{\scriptstyle
 = \sum_q  \alpha \frac{e_q^2}{2 \pi}  \int dz  dk_T^2 \frac{[z^2 + (1 - z)
^2] k_T^2 + m_q^2}{(\overline{Q}^2 + k_T^2)^2}
 N_c  \sigma_{q \overline{q} + p} (k_T^2)}
\ee
where the number of colours $N_c = 3$, and $e_q$ is the charge of                                                   
the quark in units of $e$.
                                                   
To determine $F_2 (x, Q^2)$ at low $Q^2$ we have to
evaluate the contributions to                                                   
$\sigma_T$ coming from the various kinematic domains.
First the contribution from the perturbative domain
with \mbox{$M^2 > Q_0^2$} and large $k_T^2$, and second                                                    
from the non-perturbative or long-distance domains.

\begin{figure}[htb]
\hspace*{2.0cm}
\centerline{\epsfig{file=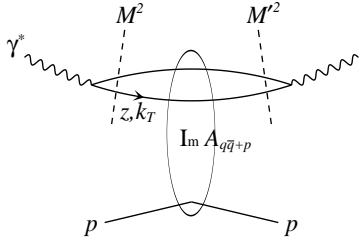,height=5cm,width=10cm}}
\vspace*{-1.0cm}
\caption{The schematic representation of the ouble dispersion \ref{eq:a1}
for the $\gamma^* p $ total cross section. The cut variables $M$ and $M'$ are the
invariant masses of the incoming and outgoing $ q \bar q $ states.}
\label{fig:fig1}
\end{figure}
\subsection{The $\gamma^* p$ cross section in the                                                   
perturbative domain}

We have to include two graphs, one shown on Fig.2 and the diagonal one, 
 for our calculation of the cross section
in the perturbative domain. Our formula for the cross section
has the following form:
\be
\label{eq:a3}
\sigma_{T,L}=F_{T,L}(z,k_{1T},l_T,Q^2) \otimes f(x,l_T^2)
\ee
where 
\be
\label{eq:a4}
f(x,l_T^2)=x \partial g (x, l_T^2) / \partial \ln  l_T^2    
\ee
is the unintegrated gluon distribution function.
The $\otimes$  denotes the convolution in the quark ($z$,$k_{1T}$)
 and gluon ( $l_T$ ) momenta variables.
The $F_{T,L}(z,k_{1T},l_T,Q^2)$ is the photon-gluon impact factor
which can be calculated perturbatively.
From the formal point of view the integrals over $l_T^2$ and                                                   
$k_T^2$ cover the interval 0 to $\infty$.  For the $l_T^2$                                                   
integration in the domain $l_T^2 < l_0^2 \sim 1 {\rm GeV}^2$ we                                                   
may use the approximation                                                   
\be                                                   
\label{eq:a5}                                                   
\alpha_S (l_T^2) \: f (x, l_T^2) \; = \; \frac{l_T^2}{l_0^2} \: \alpha_S (l_0^2)                                                   
\; f (x, l_0^2).                                                   
\ee            
The $f(x,l_0^2)$ is the input gluon distribution with free parameters
which can be adjusted to fit the data.

\subsection{Calculating the gluon distribution}
                                                  
To calculate the perturbative contributions we need to know the unintegrated gluon                                                   
distribution $f (x, l_T^2)$.  To determine it
 we carry out the full programme described in detail in                                                   
Ref.~\cite{KMS}.  We solve a \lq\lq unified\rq\rq~equation for $f (x, l_T^2)$                                                   
which incorporates BFKL and DGLAP evolution                                                  
on an equal footing, and allows the description of both small and large $x$ data.  To                                                  
be precise we solve a coupled pair of integral equations for the gluon and sea quark                                                   
distributions, as well as allowing for the effects of valence quarks. 

Schematically we can write these equations in the following form:
\bea
\label{eq:a6}
f=f^0 + \overbrace{K_{\scriptscriptstyle BFKL} \otimes f}^{\mathrm{BFKL \; part}} + 
\overbrace{(P_{gg}-1) \otimes f}^{\mathrm{DGLAP \; part}} + P_{qg} \otimes S \nonumber 
\eea
\bea
S=S^0+\underbrace{S_{\scriptscriptstyle BOX} \otimes f}_{\mathrm{k_T \; factorisation}} + P_{qq} \otimes S 
\eea

Following Ref.~\cite{KMS} we appropriately constrain                                                   
the transverse momenta of the emitted gluons along the BFKL ladder.
  There is an indication, from comparing the size of the next-to-leading 
$\ln (1/x)$ contribution                                                  
\cite{CCFL} to the BFKL intercept with the effect due to the kinematic constraint                                                  
\cite{KMS1}, that the incorporation of the constraint into the evolution analysis gives                                                  
a major part of the subleading $\ln (1/x)$ corrections.

\begin{figure}[htb]
\hspace*{7cm}
\centerline{\epsfig{file=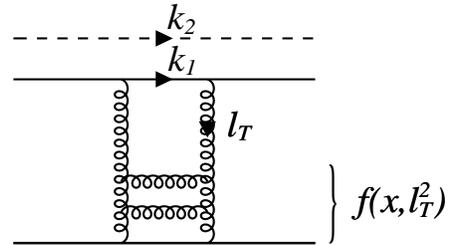,height=7cm,width=20cm}}
\vspace*{-3cm}
\caption{The quark-proton interaction via two gluon exchange}
\label{fig:fig2}
\end{figure}
 As in                               
Ref.~\cite{KMS} we take $l_0^2 = 1$~GeV$^2$, but due to the large anomalous                               
dimension of the gluon the results are quite insensitive to the choice of $l_0$ in the                               
interval 0.8--1.5~GeV.                              
                                                  
The starting distributions for the evolution are specified in terms of three parameters                                                   
$N, \lambda$ and $\beta$ of the gluon                                                  
\be                                                  
\label{eq:a7}                                                  
f_{0} (x, l_0^2) \; = \; N x^{- \lambda} (1 - x)^\beta                                                  
\ee                                                  
where $l_0^2 = 1$~GeV$^2$. 
The input for the quark sea has the following form:
\be                                                  
\label{eq:a8}                                                  
S_u^{\rm np} \; = \; S_d^{\rm np} \; = \; 2S_s^{\rm np} \; = \; C \: x^{-0.08} (1 -                                                   
x)^8.                                                  
\ee

\section{The $\gamma^* p$ cross section in the                                                   
non-perturbative domain}

There are two different non-perturbative contributions.
For $M^2<Q_0^2$ we use the conventional vector
meson formulae whereas for $M^2>Q_0^2$ and $k_T^2<k_0^2$
we use the additive quark model and the impulse approximation.

\subsection{Vector meson dominance part}

As was already mentioned we assume the vector meson
dominance model to be valid in the region where
$M^2<Q_0^2$, and $Q_0^2 \sim 1.5 \; GeV^2$.
We include three resonances: $\rho, \omega, \phi $.
For completness we should also include longitudinal
structure function $F_L(x,Q^2)$.
$F_L$ is                                                   
given by a formula just like (\ref{eq:a7}) but with the                                                   
introduction of an extra                                                   
factor $\xi Q^2/M_V^2$ on the right-hand side.  $\xi (Q^2)$ is a                                                   
phenomenological function which should decrease with increasing                                                   
$Q^2$.  The data for $\rho$ production indicate that 
$\xi (m_\rho^2) \lapproxeq 0.7$ \cite{EE}, whereas at large $Q^2$ the                                                   
usual properties of deep inelastic scattering predict that                                                    
\be                                                   
\label{eq:a9}                                                   
\frac{F_L}{F_T} \; \sim \; \frac{4 k_T^2}{Q^2} \; \lapproxeq \;                                                   
\frac{M_V^2}{Q^2}.                                                   
\ee                                                   
So throughout the whole $Q^2$ region the contribution of $F_L$ is                                                   
less than that of $F_T$.  In order to calculate $F_L$ (VDM) we                                                   
multiply the VDM prescription for $F_T$ with the factor $\xi Q^2 / M_V^2$  and use an                                                   
interpolating formula for $\xi$                                                   
\be                                                   
\label{eq:a10}                                                   
\xi \; = \; \xi_0 \left(\frac{M_V^2}{M_V^2 + Q^2} \right)^2                                                   
\ee                                                   
with $\xi_0 = 0.7$.

\subsection{Additive quark model and the impulse approximation}

For large produced masses, $M^2>Q_0^2$ but low quark momenta,
 $k_T^2<k_0^2$ we use the additive quark model and the impulse approximation.
Our formulae have the following form:
\bea                                                   
\label{eq:a11}
\scriptstyle {                                                   
\sigma_T^{\rm AQM} =  \alpha \sum_q  \frac{e_q^2}{2 \pi}  \int 
d z dk_T^2 \frac{[z^2 + (1 - z)^2] k_T^2 + m_q^2}{(\tilde{Q}^2 + k_T^2)^2}
N_c \sigma_{q \overline{q} + p}  (W^2)}
\nonumber
\eea                                                   
\be                                                   
\label{eq:a12}
\scriptstyle{                                                   
\sigma_L^{\rm AQM} = \alpha \sum_q  \frac{e_q^2}{2 \pi}  \int 
d z d k_T^2  \frac{4 Q^2  z^2 (1 - z)^2}{(\tilde{Q}^2 + k_T^2)^2}  N_c
\sigma_{q \overline{q} + p} (W^2)}
\ee                                                   
where for $\sigma_{q \overline{q} + p}$ we take, for the light quarks,                                                   
\be                                                   
\label{eq:a13}                                                   
\sigma_{q \overline{q} + p} \: (W^2) \; = \; \frac{2}{3} \; \sigma_{p                                                    
\overline{p}} \: ( s = \textstyle{\frac{3}{2}} W^2).                                                   
\ee                                                   

To allow for the confinement we replaced 
$\overline{Q}^2$ by $\tilde{Q}^2                                                    
= \overline{Q}^2 + \mu^2$ in (\ref{eq:a12}), where $\mu$ is                                                    
typically the inverse pion radius.  We therefore take $\mu^2 = 0.1                                                    
{\rm GeV}^2$.  This change has no effect for $Q^2 \gg \mu^2$ but for $Q^2                                                    
\lapproxeq \mu^2$ it gives some suppression of the AQM contribution.                                             

\subsection{The quark mass}

In the perturbative QCD domain we use the (small) current quark                                                   
mass $m_{{\rm curr}}$, while for the long distance contributions                                                   
it is more natural to use the constituent quark mass $M_0$.  To                                                   
provide a smooth transition between these values (in both the AQM and perturbative                               
QCD domains) we take the running mass obtained                                                    
from a QCD-motivated model of the spontaneous chiral symmetry breaking in the                                                    
instanton vacuum \cite{INST}                                                   
\be                                                   
\label{eq:a14}                                                   
m_q^2 \; = \; M_0^2 \: \left ( \frac{\Lambda^2}{\Lambda^2 + 2 \mu^2} \right )^6 \: +                                                    
\: m_{\rm curr.}^2.                                                   
\ee                                                   
The parameter $\Lambda = 6^{1/3}/\rho = 1.09$~GeV, where $\rho = 1/(0.6~{\rm                                                    
GeV})$ is the typical size of the instanton.  $\mu$ is the natural scale of the problem,                                                    
that is $\mu^2 = z (1-z) Q^2 + k_T^2$ or $\mu^2 = z (1-z) Q^2 + (\mbox{\boldmath                                                    
$l$}_T + \mbox{\boldmath $k$}_T)^2$ as appropriate.  For constituent and current                               
quark masses we take $M_0 = 0.35$~GeV and $m_{\rm curr} = 0$ for the $u$ and                               
$d$ quarks, and $M_0 = 0.5$~GeV and $m_{\rm curr} = 0.15$~GeV for the $s$                               
quarks.                                                     
  
\section{Summary}

We have made a fit of the $\sigma_{\gamma^{*}p}$ over the entire range
of $Q^2$ values. It relies only on the form of the initial gluon
distribution see \ref{eq:a7} and the boundary between perturbative
and non-perturbative contributions.
The advantage of our treatment is that we use a full set of
integro-differential BFKL and DGLAP equations. These equations are valid over the entire perturbative region. We have used very few free parameters which
are used to parametrise the non-perturbative region. We use VDM which is well established in this region. 
We have used running mass prescription in our calculation. The growth
of $m_q$ in the transition region is an important non-perturbative effect
which we find is required by the $F_2$ data. 

The full and complete calculation of the cross section together
with the longitudinal part is given in \cite{MRS}.
\begin{figure}[htb]
\centerline{\epsfig{file=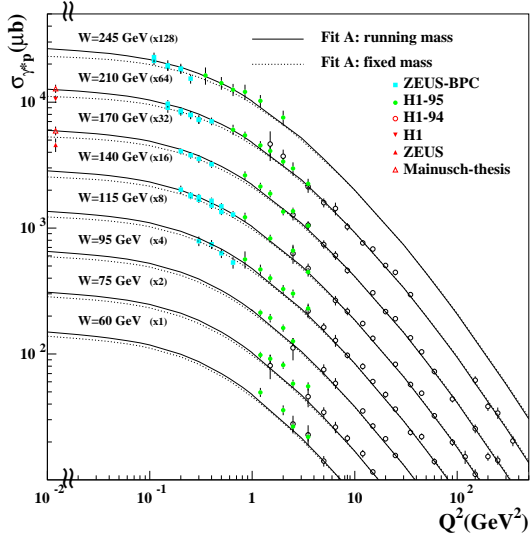,height=8cm,width=8cm}}
\caption{The curves show the virtual photon-proton cross section as a function of $Q^2$ for various values of the energy $W$}
\label{fig:fig3}
\end{figure}

\noindent {\bf Acknowledgements}
MGR thanks the Royal Society, INTAS (95-311) and the Russian Fund of                                
Fundamental                                                    
Research (98~02~17629), for support.  AMS thanks the Polish State Committee for                                             
Scientific Research (KBN) grants No.~2 P03B~089~13 and 2~P03B~137~14 for                                             
support.  Also this work was supported in part by the EU Fourth Framework                                             
Programme \lq Training and Mobility of Researchers', Network \lq Quantum                                             
Chromodynamics and the Deep Structure of Elementary Particles', contract                                             
FMRX-CT98-0194 (DG~12 - MIHT).

\end{document}